\documentclass[twocolumn,superscriptaddress,amssymb,amsmath,nobibnotes,aps,prl]{revtex4}

\usepackage{color}
\usepackage{graphicx}
\usepackage{comment}

\newcommand{\Tr}{\operatorname{Tr}}

\newcommand{\F}{{\bf{F}}}
\newcommand{\G}{{\bf{G}}}

\newcommand{\kk}{{\bf{k}}}
\newcommand{\xSigma}{{\bf{\Sigma}}}

\newcommand{\AdS}{{\rm{AdS}_2}}
\newcommand{\gi}{{g_{\mbox{\scriptsize I}}}}
\newcommand{\gii}{{g_{\mbox{\scriptsize II}}}}

\newcommand{\iar}[1]{\textcolor{blue}{\bf [[IR: #1]]}}

\begin{document}

\title{Moving mirrors, Page curves and bulk entropies in AdS$_2$.}

\author{Ignacio A. Reyes}
\email{ireyesraffo@gmail.com}
\affiliation{Max-Planck-Institut f\"ur Gravitationsphysik, Am M\"uhlenberg 1, 14476
  Potsdam, Germany}

\date{\today}

\begin{abstract}
Understanding the entanglement of radiation in QFT has been a long standing challenge, with implications ranging from black hole thermodynamics to quantum information. 
We demonstrate how the case of the free fermion in $1+1$ dimensions reveals the details of the density matrix of the radiation produced by a moving mirror. 
Using the resolvent method rather than standard CFT techniques we derive the R\'enyi entropies, modular Hamiltonian and flow of the radiation, and determine when mirrors generate unitary transformations. 
\end{abstract}

\maketitle

\section{Introduction}


It is well known that the physics of moving mirrors in QFT is intimately connected -- and in some cases equivalent -- to the thermodynamics of black holes. This relation has proven very fruitful, since the former does not require involved geometric considerations but rather only some fundamental notions about quantum fields.  

Two main strategies have been traditionally used in this subject. The first one consists of studying the global properties of the asymptotic state and resembles more closely Hawking's original calculation\,\cite{Hawking:1974rv,Davies:1976hi,Davies:1977yv,Ford:1982ct,Chen:2017lum,Good:2019tnf,Good:2016atu,Hotta:1994ha}. More recently techniques coming from gauge/gravity duality, specifically the Ryu-Takayanagi formula and its generalisations\,\cite{Ryu:2006bv,Hubeny:2007xt,Faulkner:2013ana}, allow to translate the problem into gravitational physics in a higher dimensional space\,\cite{Almheiri:2019psf,Almheiri:2019hni,Akal:2020twv}. This second approach relies strongly on methods of conformal field theory. And despite significant progress, some important questions remain open.

Often the above approaches restrict to studies of entanglement entropy. However, a quantum state is not determined only by its entanglement entropy. What one would like to figure out is the structure of the density matrix itself as the system evolves. This means that we seek to understand the correlations between arbitrary \textit{subsystems} of the radiation, a property that is neither global nor fixed by conformal symmetry. In this work we do precisely this for a very simple system: the chiral fermion in $1+1-$dimensions with a reflecting boundary. The advantage is that here we have the luxury of using the method of the resolvent\,\cite{Casini:2009vk}. 

Finding the R\'enyi entropies is particularly important in connection with the information paradox. Indeed, unitarity requires that not only the von Neumann entropy but \textit{all} R\'enyi entropies follow a Page curve. We will see under which conditions this is true for moving mirrors, and quantify the correlations between the early and late radiations. A key ingredient will be the entanglement between the two chiralities created by the mirror. 


We begin by specifying the physical system and stating the questions we wish to address.

\section{Fermions and mirrors.}

We consider the standard massless Dirac action over a patch $\mathcal{M}$ of $1+1$-dimensional Minkowski spacetime. As usual, using lightcone coordinates $x^\pm= t\pm x $ this reads
\begin{align}\label{eq:action}
I&=\frac{i}{2}\int_{\mathcal{M}} dxdt\ \left( \psi^\dag_- \partial_+ \psi_- + \psi^\dag_+ \partial_- \psi_+\right)\,.
\end{align}

We are interested in the case when $\mathcal{M}$ has a boundary $\partial \mathcal{M}$ along a worldline specified by a differentiable monotonically increasing function
\begin{align}\label{eq:mirrors}
x^+=g(x^-)
\end{align}
and we will demand $g'>0$ so that its trajectory is causal. For definiteness, we choose the physical region to be that on the \textit{right} of the boundary, so that incoming (outgoing) modes correspond to left (right) movers $\psi_+$ ($\psi_-$). 

In order for any equations of motion to follow from an action principle, we must require that the action has an extremum. Upon variation, the action gives $\delta I=\int_{\mathcal{M}}  \text{e.o.m.} + B$.
The first term involves the equations of motion so it vanishes whenever $\partial_\pm \psi_\mp=0$, while the second term is a total derivative  
\begin{align}\label{}
B=\frac{i}{2}\int_{\partial \mathcal{M}} dx^- \left( \psi_-^\dag \delta \psi_- - g'(x^-) \psi_+^\dag \delta \psi_+  \right)+\text{h.c.}
\end{align}
and must be required to vanish. 
A natural condition that achieves $B=0$ is that $\partial\mathcal{M}$ acts as a `mirror' by imposing reflecting boundary conditions,
\begin{align}\label{eq:bc}
\psi_-(x^-)=\epsilon \sqrt{g'(x^-)} \psi_+\left( g(x^-) \right)\,.
\end{align}
Incoming modes reaching the boundary (mirror) are reflected as outgoing modes, the choice $\epsilon=\pm 1$ corresponding to whether the wave flips orientation upon reflection. 

\begin{figure}[h]
\includegraphics[width=0.3\textwidth]{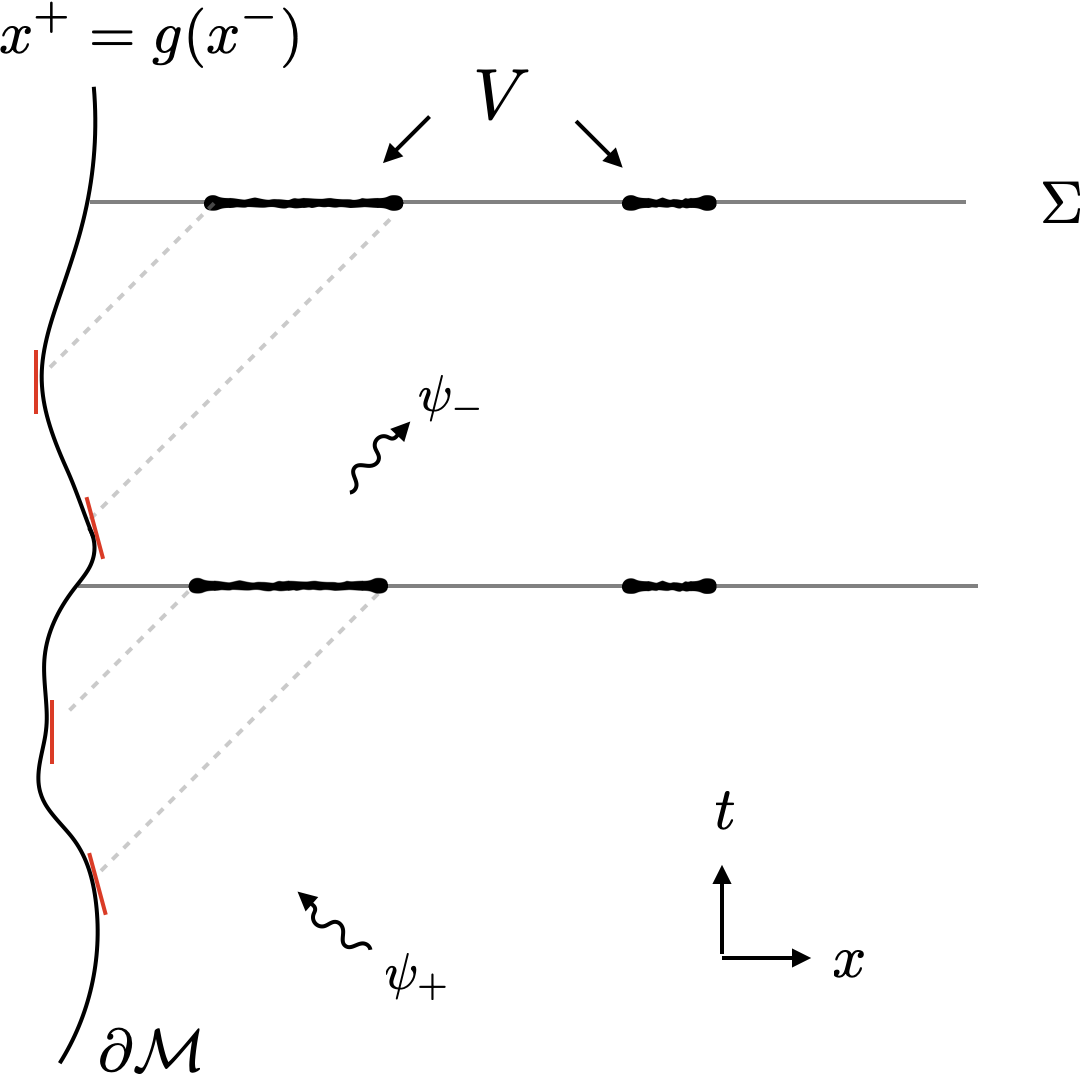}
\caption{Evolution of an entangling region $V$ for a given mirror trajectory. The R\'enyi entropies of $V$ depend on the mirror only through the position and velocity at the null projections of the region's boundary, as examplified in \eqref{eq:Omegainterval}. Thus if these return to their original values, the mirror produces unitary transformations. }
\label{fig:general}
\end{figure}

We will not consider standard boundary conformal field theory (BCFT)\,\cite{DiFrancesco:1997nk} because arbitrary mirror trajectories in general break all conformal symmetries. In other words, the boundary condition \eqref{eq:bc} are not `conformal boundary conditions': although the stress tensor remains traceless due to the equations of motion, its parallel/perpendicular components -- which measures the energy flowing \textit{away} at the boundary -- does not vanish but is governed by the anomaly. If the incoming state is the vacuum, this gives
\begin{align}\label{eq:Tparper}
\langle :T_{\parallel\perp}:\rangle\big|_{\text{mirror}} = \frac{\left( Sg \right)(z)}{12\pi g'(z)} 
\end{align}
where $Sg$ stands for the Schwarzian derivative. Thus energy will be injected/extracted to/from the system by the moving mirror, similar to what the gravitational field does to the quantum fields outside the horizon.  
As initial data, we must provide the quantum state on some Cauchy surface. 
For simplicity we focus on mirrors that are asymptotically static in the past, so that the incoming state is well defined along past null infinity $\mathcal{I}^-_R$, see fig. \ref{fig:hawking}. 



In a free theory, the two-point function plays a major role. Throughout the text we denote by
\begin{align}\label{eq:G}
G(x,y)\equiv\langle \psi_+(x)\psi_+^\dag(y)\rangle
\end{align}
the \textit{incoming/incoming} (left/left) correlation function, i.e. the initial data on $\mathcal{I}^-_R$. Here $x,y$ are spatial coordinates along a Cauchy slice as described below. 
Although our analysis is valid for a larger class of gaussian states, we focus on incoming equilibrium states. This simplifies the discussion and emphasises the role of the mirror rather than the initial data. An incoming state prepared on $\mathcal{I}^-_R$ at inverse temperature $\beta$ is given by
\begin{align}\label{eq:Gthermal}
G(x,y)=\frac{1}{2i\beta \sinh\left( \pi(x-y)/\beta \right)}\ .
\end{align}

Because both chiralities are involved, we must consider the correlation matrix for the Dirac spinor $\Psi=(\psi_+,\psi_-)$,
\begin{align}\label{eq:Gbold}
{\bf{G}}(x_1,x_2)&=\langle \Psi(x_1)\Psi^\dag(x_2)\rangle=\left( \begin{array}{cc} G_{++} & G_{+-} \\ G_{-+} & G_{--} \end{array}  \right)
\end{align}
where $G_{ij}\equiv\langle \psi_i \psi^\dag_j\rangle$ with $i,j=\pm $. Two-dimensional spinor matrices are denoted in boldface. While $G_{++}$ is initial data, the boundary conditions \eqref{eq:bc} determine the remaining entries of the correlator in terms of $G$:
\begin{align}\label{eq:Gij}
G_{ij}
&=\epsilon^{\frac{i-j}{2}} \sqrt{ (-1)^{\frac{i-j}{2}} q'_i(x) q'_j(y)} \  G(q_i(x),q_j(y))
\end{align}
where we have defined $q_+(x)\equiv x^+,q_-(x)\equiv g(x^-)$.


We now choose an incoming state specificied by \eqref{eq:Gthermal} and a mirror trajectory specified by $g(\cdot)$. Together these determine a global state $\rho_\Sigma$ on each Cauchy slice $\Sigma$, see fig. \ref{fig:general}. We now choose a subregion $V\subset \Sigma$ consisting of $N$ disjoint intervals $V=\cup_\ell (a_\ell,b_\ell)$ with $\ell=1,\hdots, N$, and wish to compute the reduced density matrix
\begin{align}\label{}
\rho_V=\Tr_{V^c}\left( \rho_\Sigma \right)
\end{align}
 obtained by tracing out the complement of $V$ along $\Sigma$. For simplicity we restrict to Cauchy slices of constant time, the generalisation being straightforward. \footnote{In the continuum limit the existence of the density operator is of course problematic. In this paper we abuse notation and use this notation since it is the more familiar language.   }

Here is where having a free theory comes in handy\,\cite{Peschel_2003}. For free fermions, it is sufficient that $\rho_V$ reproduces the correlator $\G(x,y)=\text{tr}_V\left( \rho_V \Psi(x) \Psi^\dag(y) \right)$ for $x,y\in V$.
Now any state can be written as $\rho_V= e^{-K_V}$
where $K$ is called the \textit{modular Hamiltonian}. For gaussian states this takes a quadratic form
\begin{align}\label{eq:Kmod}
K=\int_{V^2}dxdy\ \Psi^\dag(x) \kk(x,y)  \Psi(y)
\end{align}
with kernel\,\cite{Peschel_2003} 
\begin{align}\label{eq:kernel}
\kk=-\log \left( \G^{-1}-\bf{1} \right)\,.
\end{align}
Here and below $\G(x,y)$ is taken as a linear operator acting on smooth functions supported on $V$ via convolution. 

This translates the problem of finding the reduced density matrix into that of computing functions of $\G$ on $V$, but with both chiralities simultaneously. This sets the stage for the application of the resolvent method, a technical tool that is reviewed in the Supplemental Material. The basic idea is that in order to compute functions of an operator one can use Cauchy's integral formula,
\begin{align}\label{}
f(\G)=\frac{1}{2\pi i} \oint_\gamma d\lambda \frac{f(\lambda)}{\lambda - \G}\,,
\end{align}
where $\gamma$ encircles the spectrum of $\G$. The operator $\left( \lambda-\G \right)^{-1}$ is called the resolvent of $\G$. 

The simplest application of this method is to compute the entropies, to which we now turn. 


\section{Entropies}

The entanglement R\'enyi entropies are defined as
\begin{align}\label{}
S^{(n)}=\frac{1}{1-n}\log \frac{\Tr\left( \rho^n \right)}{\left( \Tr \rho \right)^n}\,.
\end{align}

For the free fermion, it is easy to show that
\begin{align}\label{}
\log \Tr\left( \rho^n \right)=\Tr \log \left( \G^n+(1-\G)^n \right)\,.
\end{align}
In the Supplemental Material we show how to use the resolvent to compute these expressions. After the dust settles, the final result for the R\'enyi entropies is
\begin{align}\label{eq:Snmirror}
S^{(n)}
=\frac{n+1}{24n} \log \frac{\Omega(x^+)}{\Omega(g(x^-))}  \Big|_{\partial V_\delta}\,,
\end{align}
where
\begin{align}\label{eq:Omegasol}
\Omega(x)
&=-\prod_{\ell=1}^N \frac{G(x,b_\ell^+)}{G(x,a_\ell^+)} \frac{G(x,g(a_\ell^-))}{G(x,g(b_\ell^-))}\,
\end{align}
and we have introduced the region $V_\delta=\cup_{\ell}(a_\ell+\delta,b_\ell-\delta)$ with a very small $\delta>0$ in order to regularise the UV divergences. 
Throughout the paper all $n-$dependence of R\'enyi entropies appears as an overall factor. 


Here it is illustrative to look at a specific example. Consider the vacuum as incoming state and a single interval $(a,b)$ at time $t$ but leaving the mirror trajectory arbitrary. The entropies \eqref{eq:Snmirror} yield
\begin{align}\label{eq:Omegainterval}
S^{(n)}
&=\frac{n+1}{12n} \log  \left( \frac{b-a}{\delta^2} \frac{\frac{b^+-g(b^-)}{b^+-g(a^-)}}{\frac{a^+-g(b^-)}{a^+-g(a^-)}} \frac{g(a^-)-g(b^-)}{ \sqrt{g'(b^-)g'(a^-)}}  \right)\,.
\end{align}

This result is remarkably simple. It depends on the mirror position $g$ and velocity $g'$ only at $a^-,b^-$, i.e. where the null projections of $\partial V$ intersect the mirror trajectory, see figure \ref{fig:general}.

This gives rise to a unitarity criterion in the following way. Consider a fixed interval $(a,b)$ together with a mirror trajectory $g(t)$. 
\textit{If the position and velocity of the mirror at the null projections of $\partial V$ are identical at $t_1$ and $t_2$, then all entropies \eqref{eq:Omegainterval} at $t_1$ are equal to those at $t_2$, and the mirror generates states related by a unitary transformation}
\begin{align}\label{}
U\rho_V[g(t_1)]\,U^\dag=\rho_V[g(t_2)]\,.
\end{align}
The behaviour of the mirror anywhere else is irrelevant. This case is depicted in figure \ref{fig:general}. 
 Here it is crucial to observe that, even if all the R\'enyi entropies coincide, the \textit{states} $\rho_V[g(t_{1,2})]$ can be very different. 
 Indeed, gaussian states are determined by their correlation functions, and two different mirrors with identical $g(a^-),g(b^-),g'(a^-),g'(b^-)$ produce the same entropies but different correlators. 

A useful tool to quantify unitarity questions is the mutual information, as we show next.



\subsection{Mutual Information}

Mutual information (MI) is defined as $I(V_1|V_2)=S^{(1)}(V_1)+S^{(1)}(V_2)-S^{(1)}(V_1\cup V_2)$. It measures the inherent correlations between $V_1$ and $V_2$ and is a better quantifier of correlations than entropy since it is UV finite. 
From \eqref{eq:Snmirror} we find that the mutual information in the presence of a moving mirror decomposes as
\begin{align}\label{eq:mutual}
I=I_{\text{plane}}+\frac{1}{6}\log \omega\,.
\end{align}
The first term is the usual mutual information of two independent chiralities on the plane (no boundary), while the second term is the new contribution due to the mirror. To illustrate them, let us consider the simple case of the vacuum as incoming state. Then, we have as usual $I_{\text{plane}}=\frac{1}{6}\log \frac{(a_2-a_1)(b_2-b_1)}{(b_2-a_1)(a_2-b_1)}$.  
The novel term due to the mirror is given by
\begin{align}\label{eq:omega}
\omega=\frac{  \frac{\left(b_2^+-g(a_1^-)\right)  \left(b_1^+-g(a_2^-)\right) \left( g(b_2^-)-g(b_1^-) \right)     }{\left( b_2^+-g(b_1^-) \right) \left( b_1^+ - g(b_2^-) \right) \left( g(b_2^-)-g(a_1^-) \right)  }         }{     \frac{  \left( a_2^+-g(a_1^-) \right) \left( a_1^+-g(a_2^-) \right)   \left( g(a_2^-)-g(b_1^-) \right)       }{   \left(a_2^+-g(b_1^-)\right)  \left(a_1^+-g(b_2^-)\right)   \left(g(a_1^-)-g(a_2^-)\right)    }        }\,,
\end{align}
It is noteworthy that although the individual entropies \eqref{eq:Omegainterval} depend on the mirror velocity $g'$, mutual information is independent of it. The  velocity dependence arises via the UV divergences, which mutual information is devoid from by construction. 

For example, for the static mirror, \eqref{eq:omega} is given by
\begin{align}\label{eq:wRHP}
\omega_{\text{RHP}}=\frac{\sinh \frac{\pi(a_1+b_2)}{\beta}\sinh \frac{\pi(a_2+b_1)}{\beta}}{\sinh \frac{\pi(a_1+a_2)}{\beta} \sinh \frac{\pi(b_1+b_2)}{\beta}    } .
\end{align} 
In the examples below, mutual information will prove useful to quantify the violation of unitarity produced by different mirrors. Having understood the entropies, we now move on to another important aspect dealing with the action of the density matrix itself.

\section{Modular Hamiltonian and Flow}
\label{sec:Modular}


The modular Hamiltonian \eqref{eq:Kmod} defines the \textit{modular flow}, an automorphism of the algebra of observables supported on the causal region associated to $V$. For any operator $\mathcal{O}$, it is defined by evolving with the modular Hamiltonian
\begin{align}\label{eq:modflow}
\sigma_\tau\left( \mathcal{O} \right)\equiv e^{-i\tau K} \mathcal{O} e^{i\tau K}\,.
\end{align}
with $\tau$ the `modular time'. The simplest flow to study is that of the fundamental field itself. Using the tools in \cite{Erdmenger:2020nop}, we find that
\begin{align}\label{eq:sigmat}
\sigma_\tau\left( \psi_i (x) \right)=\int_V dy\, \Sigma_{ij}(x,y)\psi_j(y) 
\end{align}
with the modular kernel $\xSigma=\left( \G^{-1}-1 \right)^{-i\tau}$. 
 And once more the resolvent allows to computes this, the result is
\begin{align}\label{eq:Sigmaijsol}
\Sigma_{ij}
=2\pi i\sinh(\pi \tau)\,\delta \left( Z(q_i(x))-Z(q_j(x)) -\tau\right) G_{ij}(x,y).
\end{align}

To understand the locality properties of the flow, we must examine the number and nature of the solutions to
\begin{align}\label{eq:ZZ}
Z(q_i(x))-Z(q_j(y))-\tau=0\,.
\end{align}

For our purposes only one property of the function $Z(\cdot)$ is relevant: $Z(q_i)$ for $i=\pm$ increases/decreases monotonically from $\pm\infty$ to $\mp\infty$ in each interval of $V$. 
Therefore, there exists a unique solution to \eqref{eq:ZZ} in each interval as follows. For equal chiralities, $i=j$, we call the solutions $y_\ell(x)$, where $\ell$ labels the interval. These are similar to those already encountered in \cite{Casini:2009vk,Klich:2017qmt,Blanco:2019xwi,Erdmenger:2020nop}, and includes the \textit{local} solution $y=x$. For opposite chiralities $i=- j$, we have a novel set of solutions which we call $y_{-\ell}$ to indicate that they are associated to a change in chirality. 

Introducting the kernel \eqref{eq:Sigmaijsol} into \eqref{eq:sigmat} yields the explicit form of the modular flow:
\begin{align}\label{eq:sigmatfull}
\sigma_\tau\left( \psi_i(x) \right)&=2\pi \sinh(\pi \tau) \sum_\ell \left[ \frac{G_{ii}(x,y_\ell(x)) \psi_i(y_\ell(x))}{|\partial_y Z(y)\, \partial_y q_i(y)  |_{y=y_\ell(x)}} \right.\nonumber\\
&\left.+\, \frac{G_{i,-i}(x,y_{-\ell}(x)) \psi_{-i}\left( y_{-\ell}(x) \right)}{|\partial_y Z(y) \partial_y q_{-i}(y)|_{y=y_{-\ell}(x)}} \right]\,.
\end{align}

Modular flow takes the field of chirality $i$ at point $x$, and evolves it to two points on each interval: a contribution of the same chirality located at $y_{\ell}(x)$, and another with opposite chirality located at $y_{-\ell}(x)$, see fig. \ref{fig:flow}. 




After these formal developments, we turn our attention to some specific examples of mirror trajectories that are of particular physical interest.

\section{Examples of mirrors.}


\subsection{A static mirror (RHP).}

As our simplest example, consider a single interval on the right half plane (RHP), i.e. a static boundary at $x=0$. This does give rise to a conformal boundary condition where \eqref{eq:Tparper} vanishes. For the vacuum as incoming state, the R\'enyi entropies \eqref{eq:Snmirror} give
\begin{align}\label{eq:Renyistatic}
S_{\text{RHP}}^{(n)}
=\frac{n+1}{12n} \left[ \log \left( \frac{b-a}{\delta} \right)^2 + \log \frac{4r}{(r+1)^2}  \right]
\end{align}
where $r=b/a$. These agree with those reported in \cite{Mintchev:2020uom} recently. The first term is simply twice the universal entropy of a single chirality, while the second one is due to the mirror. If the incoming state is thermal, the entropies are instead
\begin{align}\label{eq:Renyistatic2}
S_{\text{RHP}}^{(n)}
&=\frac{n+1}{12n}  \left[  \log \left( \frac{\beta}{\pi \delta}  \sinh \frac{\pi(b-a)}{\beta} \right)^2 + \log \frac{4\tilde r}{(\tilde r+1)^2} \right]
\end{align}
where $\tilde r=\tanh\frac{2\pi a}{\beta} / \tanh\frac{2\pi b}{\beta}$. Notice that the second term in both \eqref{eq:Renyistatic} and \eqref{eq:Renyistatic2} is always negative so the mirror \textit{lowers} the entropy of the system. 

\begin{figure}[h]
\includegraphics[width=0.5\textwidth]{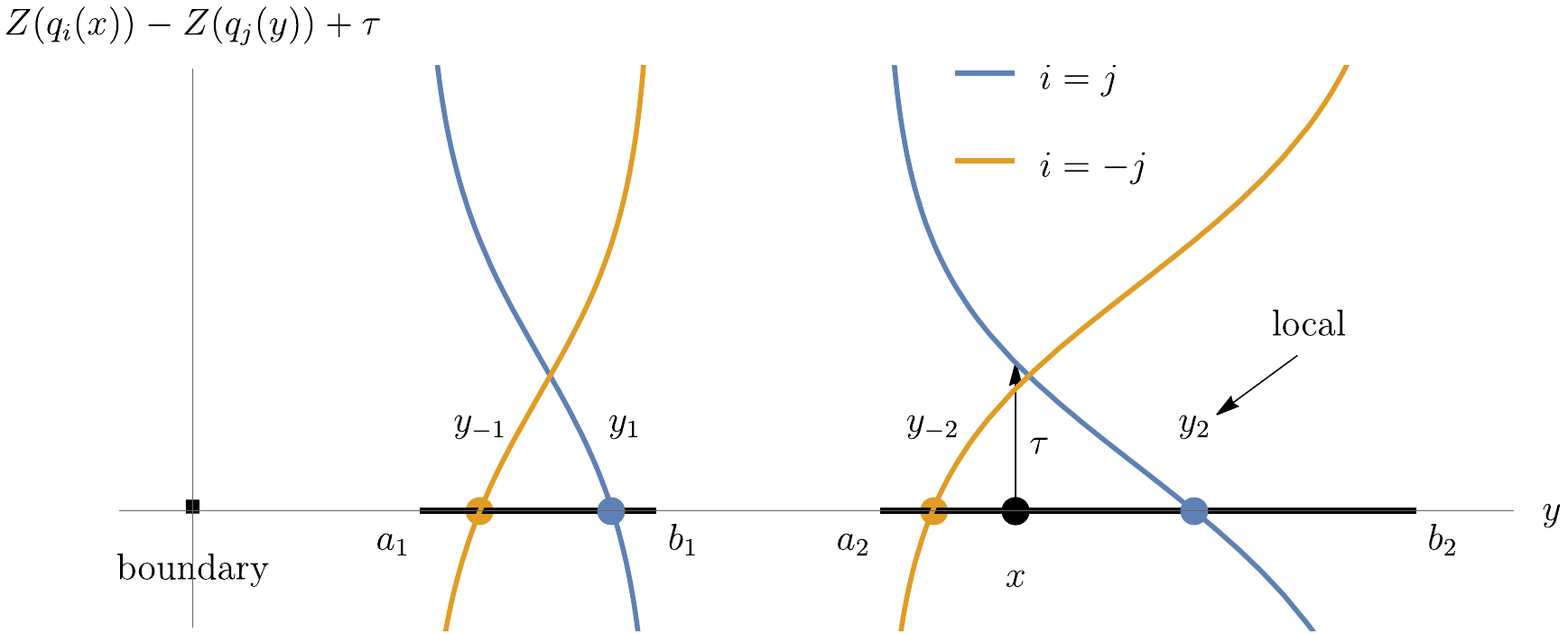}
\caption{ Illustration of the modular evolution governed by \eqref{eq:sigmat}, for two disjoint intervals and a static mirror and incoming vacuum. We plot \eqref{eq:ZZ}, for equal and opposite chiralities (this is not a spacetime diagram). The modular flow of $\psi_i(x)$ yields, at each interval, contributions of both chiralities $\psi_{\pm i}$ located at the points $y_{\pm\ell}(x)$ that are solutions to \eqref{eq:ZZ} (coloured circles). Evolution in modular time $\tau$ shifts the curves vertically, and the roots evolve accordingly. The root at $y_2$ is the local (geometric) solution, with $y_2\to x$ as $\tau\to 0$. Here $a_1=1,b_1=2,a_2=3,b_2=5.3$. }
\label{fig:flow}
\end{figure}

Moving now to the modular flow, the case of two intervals $V=(a_1,b_1)\cup (a_2,b_2)$ is depicted in fig. \ref{fig:flow}. 
As we saw above, eq. \eqref{eq:ZZ} determines which points are coupled along the flow. In this case \eqref{eq:ZZ} a quartic equation yielding four real solutions  $y_{\pm 1,2}$. For more intervals, the situation is completely analogous, involving two points per interval.  

\subsection{From vacuum to thermal and back.}

\begin{figure}[h]
\includegraphics[width=0.35\textwidth]{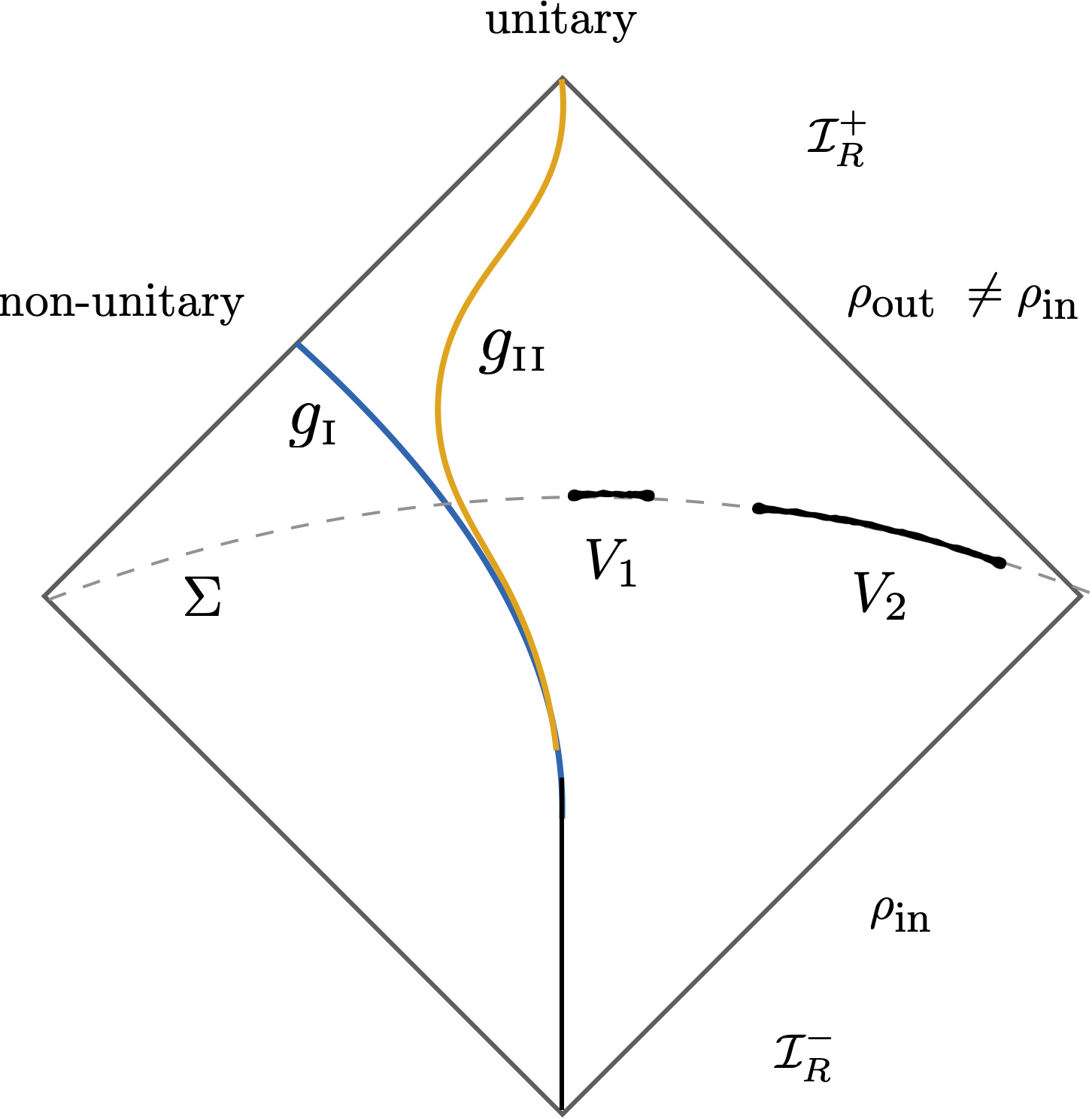}
\caption{Two different mirror trajectories with entangling region $V=V_1\cup V_2$.  Although the incoming and outgoing states $\rho_{\mbox{\scriptsize in/out}}$ are very different, unitarity means that their R\'enyi entropies on arbitrary $V$ match. Illustrated are $g_{\mbox{\tiny I}}$ and $g_{\mbox{\tiny II}}$ from the main text: the former starts to accelerate becoming asymptotically null, describing a non-unitary process. The latter follows the former for some time before returning to its original trajectory and respects unitarity. }
\label{fig:hawking}
\end{figure}

Next, we focus on an accelerating trajectory that shares the characteristic feature of Hawking radiation, namely that the outgoing state measured at $\mathcal{I}^+$ is thermal. In our moving mirror setup, there is a unique mirror profile that takes an incoming vacuum and renders the outgoing modes in a thermal state. This is
\begin{align}\label{eq:gthermal}
g_{\text{I}}(x^-)=\frac{\beta}{\pi} \tanh\left( \frac{\pi}{\beta} x^- \right)\ \ \ \mbox{\footnotesize vacuum}\to \mbox{\footnotesize thermal}\,.
\end{align}
Since this trajectory becomes null also in the remote past producing singularities, its early stage must be replace by a timelike one as we have mentioned above, see fig. \ref{fig:hawking}. As is well known\,\cite{Davies:1976hi}, this class of trajectories are closely related to the exterior region of a black hole formed by collapse, with the associated problem of information loss.  

Interestingly the converse effect is also possible: given an incoming thermal state, any trajectory that asymptotes to the inverse function, namely $\tilde g_{\text{I}}(x^-)=\frac{\beta}{\pi} \mbox{arctanh}\left( \frac{\pi}{\beta} x^- \right)$
takes a thermal state and reflects it as the vacuum. 

\subsection{Mirror with uniform acceleration.}

Our final example is a mirror moving with constant proper acceleration. The mirror stands static at $x=-R$ until $t=0$ when it begins to accelerate at constant rate away from the physical region, following the Rindler trajectory $t^2-x^2=-R^2$ corresponding to
\begin{align}\label{}
g_{\text{III}}(x^-)=-R^2/x^-\,.
\end{align}

Although qualitatively this profile is similar to $\gi$, it exhibits an important difference. 
The R\'enyi entropies for $t\gg b-R$ read
\begin{align}\label{eq:Saccel}
S^{(n)}&=\frac{n+1}{12n} \left[ \log \left( \frac{b-a}{\delta} \right)^2 + \frac{2R^2-(a+b)^2R^2+2(ab)^2}{t^4} \right]
\end{align}

This result displays an interesting feature. Notice the first term is identical to the vacuum R\'enyi entropy of \textit{two} independent chiralities, as if the mirror was not present. In the asymptotic future, the second term vanishes. Thus the original entanglement between the two chiralities, created by the static mirror, is `erased' exactly by the accelerating mirror, leaving two unentangled chiralities.  This is also seen directly by looking back at the correlation matrix itself, for in the limit $t\to\infty$
\begin{align}\label{eq:ttoinfty}
 G_{\pm\mp}\to 0 \ \ \ \ \ \text{for } g_{\mbox{\scriptsize III}} 
\end{align}
so that the correlations between left and right movers vanish. Because opposite chiralities are not entangled with each other any more on $V$, they become more entangled with the complement which has the effect of \textit{increasing} the entropy. This is a hallmark of non-unitary: the entropy in the distant future is larger than its counterpart in the remote past. 

\subsection{Page curves from Mutual Information.}

How can we capture the correlations between the early and late radiation? Consider again two fixed disjoint regions $V=V_1\cup V_2$ as depicted in fig. \ref{fig:hawking}. We will compare two mirror trajectories that remain static until $t=0$, when they begin to move. The first is $\gi$, already introduced, which scatters the vacuum into a thermal outgoing state. The second, $\gii$, follows $\gi$ for some time but then deviates from it in order to smoothly return to the static path. The precise functional form of $\gii$ is irrelevant. 

In principle one could simply track the entropies of these regions of space as they evolve in time as done above. However, this approach is not completely satisfactory. First, R\'enyi entropies are not well defined in the UV. Moreover, if we wish to keep track of \textit{all} the radiation that has escaped to infinity, we must consider an unbounded spatial subregion, which introduces yet another divergence. 

These problems are remedied by considering mutual information instead. MI is always finite: it is by construction free from UV divergences, and in addition we can safely take the limit $b_2\to\infty$ that stretches all the way to spatial infinity. Furthermore, MI has the natural physical interpretation that we seek: it measures the correlations between the early radiation (collected in $V_2$) and the late radiation (contained in $V_1$).

\begin{figure}[h]
\includegraphics[width=0.43\textwidth]{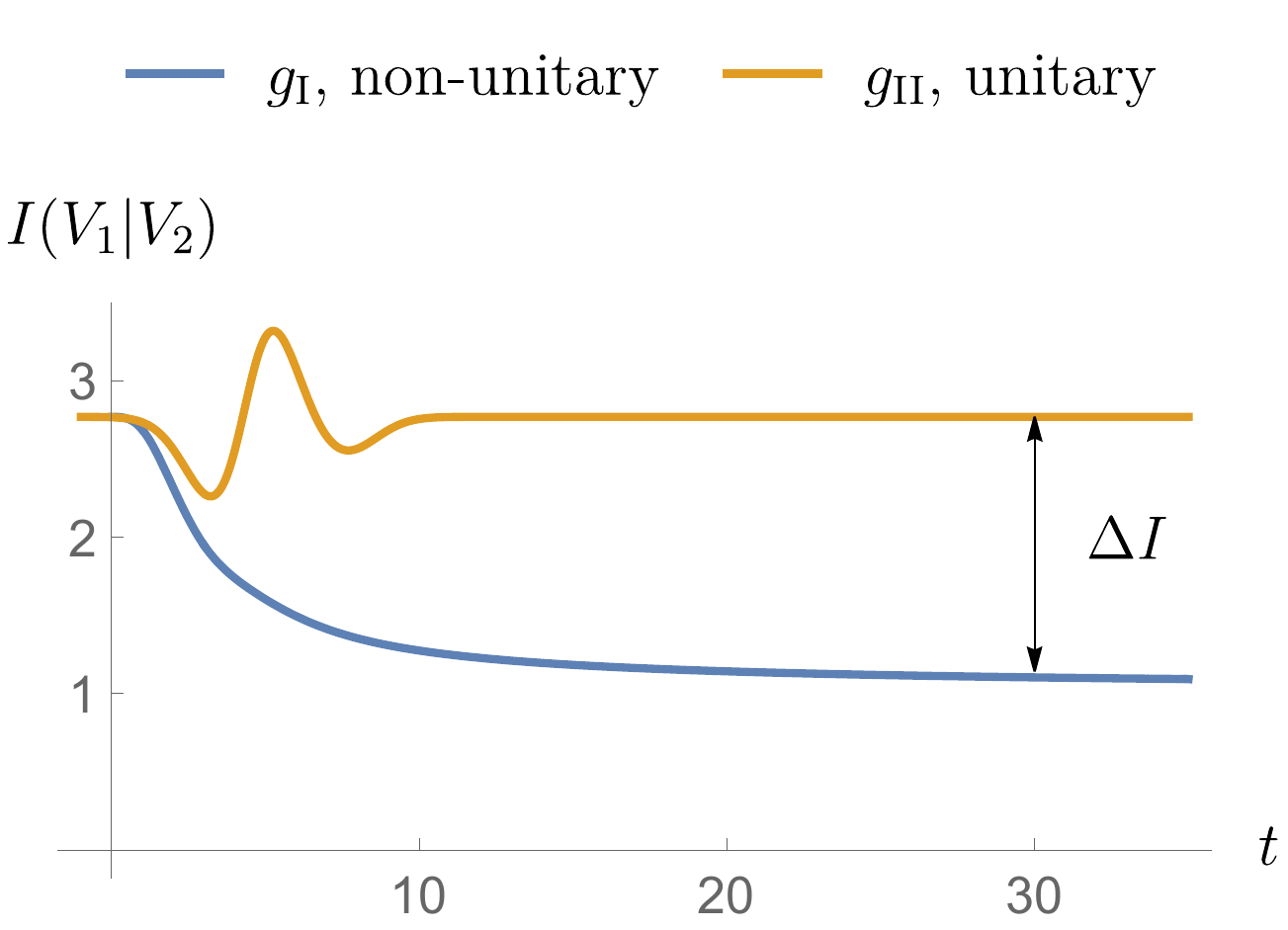}
\caption{Evolution of mutual information \eqref{eq:mutual} between the regions $V_1$ and $V_2$, for the two mirror trajectories depicted in fig. \ref{fig:hawking}, measuring the amount of correlations between the early and late radiation. MI returns to its original value for trajectory $g_{\text{II}}$, whereas it doesn't for $g_{\text{I}}$. The loss of correlations $\Delta I$ given in \eqref{eq:dI} serves to quantify unitarity violation. Here $a_1=0,b_1=1,a_2=3,b_2=10^6$ with $\beta=7$ in \eqref{eq:gthermal}. }
\label{fig:PageMI}
\end{figure}

In fig. \ref{fig:PageMI} we plot the evolution of the MI between the early and late radiation as a function of time. Clearly for the trajectory $\gi$, there is a loss of correlations between the late and early radiation compared to trajectory $\gii$, which is unitary. The asymptotic difference in MI, $\Delta I=\lim_{t\to \infty} \lim_{b_2\to \infty} \left( I_{g_{\text{II}}}(V_1|V_2)-I_{g_{\text{I}}}(V_1|V_2) \right)$, can be used to quantify the violation of unitarity as a function of the temperature of the outgoing radiation and reads
\begin{align}\label{eq:dI}
\Delta I
&=\frac{1}{3}\log \frac{(a_2-a_1)(a_2+b_1)^2\left( e^{\frac{2\pi a_2}{\beta}}-e^{\frac{2\pi b_1}{\beta}} \right)}{(a_2-b_1)(a_1+a_2)^2 \left( e^{\frac{2\pi a_2}{\beta}} - e^{\frac{2\pi a_1}{\beta}} \right)}\,.
\end{align} 
It increases monotonically until saturating at high temperature. 

\section{Conclusion}

In this paper we have investigated the entanglement in the radiation produced by a moving mirror using analytic techniques. The main physical effect of introducing a reflecting boundary is that of \textit{entangling} the two chiralities. This shows up as extra terms in the R\'enyi entropies, and in the modular flow as additional bi-local couplings due to chirality exchange. 
We found that for a static mirror, the entanglement among the chiralities always \textit{decreases} the entropies.  It would be very interesting to understand this in the context of monogamy of entanglement in QFT\,\footnote{We thank Pascal Fries for pointing out this interpretation.}.

\begin{acknowledgments}

\section{acknowledgments}

It is a pleasure to thank Ana Alonso-Serrano, Enrico Brehm, Raghu Mahajan, Juan Maldacena, Pedro F. Ramirez and the GQFI group at AEI for stimulating discussions. In particular I wish to thank Ra\'ul Arias and Rob Myers for insightful comments on an early version of the draft. Specially I would like to thank Pascal Fries for joining the initial stages of the project as well as elucidating some subtle aspects. The Gravity, Quantum Fields and Information group at AEI is generously supported by the Alexander von Humboldt Foundation and the Federal Ministry for Education and Research through the Sofja Kovalevskaja Award.

\end{acknowledgments}

\bibliography{moving_bdy_arxiv_PRL}
\bibliographystyle{ieeetr}

\newpage

\section{Supplemental Material}

\subsection{Resolvent}

This section contains the main technical tools of this work, following the approach of \cite{Fries:2019ozf,Arias:2018tmw,Erdmenger:2020nop}. Given an operator $\G$ of bounded spectrum and a function $f(\lambda)$ holomorphic in the interior of a contour $\gamma$ enclosing the spectrum -- in the case of fermions, the interval $[0,1]$ -- Cauchy's integral formula defines the function of an operator by
\begin{align}\label{eq:Cauchyformula}
f(\G)=\frac{1}{2\pi i} \oint_\gamma d\lambda \frac{f(\lambda)}{\lambda - \G}\,.
\end{align}
The challenge is of course to compute the operator $\left( \lambda-\G \right)^{-1}$ known as the \textit{resolvent}. Again, the boldface notation indicates that these are $2\times 2$ matrices associated to the Dirac spinor. The resolvent possess a branch cut along the spectrum. 

We proceed in two steps. First, we will reduce the spinorial integral equation into a more familiar one associated to a single chirality. Second, we recast the problem as a contour equation in the complex plane and solve it via residue analysis. We begin by a convenient change of variables, 
\begin{align}\label{eq:ansatz}
\frac{1}{\lambda -\G}=\frac{{\bf{1}}}{\lambda}+\frac{1}{\lambda^2} \mathbf F
\end{align}
which turns the functional equation $\left( \lambda-\G \right)\times \left( \lambda-\G \right)^{-1}=\bf{1}$ into the integral equation
\begin{align}\label{eq:SIE}
-\mathbf G(x,y)+\mathbf F(x,y|\lambda)-\frac{1}{\lambda} \int_V dz\, \mathbf G(x,z) \mathbf F(z,y|\lambda)={\bf 0}\,.
\end{align}
where the notation $\F(z,y|\lambda)$ indicates that $\F$ depends on two spacetime arguments, as well as the parameter $\lambda$ in \eqref{eq:ansatz}.


Here it is important to notice that since the mirror trajectory is subluminal, only the diagonal entries of $\bf G$ have a pole for coinciding points measured at equal times,
\begin{align}\label{eq:polediag}
G_{\pm\pm}(x\to y)&\sim \pm \frac{1}{2\pi i(x - y)}\,.
\end{align}
This shows that the presence of the mirror does not modify the UV physics, and is relevant in our analysis below. 

In order to solve \eqref{eq:SIE}, we shall first propose an Ansatz for $\F$ and transform this equation into a standard scalar integral equation. We propose a solution with a similar structure as $\bf G$ itself:
\begin{align}\label{eq:Fij}
F_{ij}(x,y|\lambda)=\epsilon^{\frac{i-j}{2}} \sqrt{ (-1)^{\frac{i-j}{2}} q'_i(x) q'_j(y)} \ F\left( q_i(x),q_j(y)|\lambda \right)\,.
\end{align}

With this Ansatz, a straightforward calculation shows that the singular integral equation becomes
\begin{align}\label{}
&-G(q_i,q_k)+F(q_i,q_k|\lambda)-\frac{1}{\lambda}\int_V dz \nonumber \\
&\left[ G(q_i,z^+)F(z^+,q_k|\lambda)+g'(z^-) G(q_i,g(z^-))F(g(z^-),q_k|\lambda)\right]=0\,. \label{eq:twochiral}
\end{align}


Since by assumption $g(\cdot)$ is smooth and monotonically increasing, \eqref{eq:twochiral} can be recasted as a simpler integral equation
\begin{align}\label{SEIscalar}
-G(x,y)+F(x,y|\lambda)-\frac{1}{\lambda}\int_{\tilde{V}} dz\, G(x,z)F(z,y|\lambda)=0
\end{align}
where now $x,y$ belong to the region $\tilde V$ defined by
\begin{align}\label{}
\tilde{V}:=V^+\cup g(V^-)
\end{align}
where
\begin{align*}
V^+=\cup_{\ell}(a_\ell^+,b_\ell^+)\ \ ,\ \ g(V^-)=\cup_{\ell}\left( g\left( b_\ell^- \right),g\left( a_\ell^- \right) \right)\,.
\end{align*}
The region $\tilde V$ consists of the null `reflections' of $V$ through the mirror. Thus, we have effectively reduced the spinorial problem on $V$ on a semi-infinite line to a scalar problem on a different region $\tilde{V}$ on the real line. 





In \cite{Fries:2019ozf} a recipe was provided to find the scalar function $F$ that solves \eqref{SEIscalar}. The first step consists of rewriting the LHS of \eqref{SEIscalar} as a contour integral in the complex plane,
\begin{align}\label{eq:contour1}
\oint_{\tilde{V}} dz\, G(x,z)F(z,y|\lambda)=0\ ,
\end{align}
while the second step is to solve the resulting equation via residues. For the first step, we require that in a neighbourhood of $\tilde{V}\subset\mathbb C$, $F$ possesses:
\begin{itemize}

\item A single pole along $\tilde{V}$
\begin{align}\label{eq:Fpoles}
F(z\to y,y|\lambda)\sim  \frac{1}{2\pi i(y-z)} +\hdots
\end{align}

\item A multiplicative branch cut along along the region,
\begin{align}\label{eq:branch}
F(z+i 0^-,y|\lambda)= \frac{\lambda-1}{\lambda} F(z-i0^+,y|\lambda)\ ,\ z\in \tilde{V}
\end{align}

\item Vanishing residues around $\partial \tilde{V}$, and analytic everywhere else in an neighbourhood of $\tilde{V}$. 

\end{itemize}

As shown in \cite{Fries:2019ozf}, these requirements are sufficient to completely determine $F$ for the class of thermal states considered in the main text. The solution to \eqref{eq:twochiral} is
\begin{align}\label{eq:Fijsol}
F_{ij}(x,z|\lambda)
&=G_{ij}(x,z)\left(\frac{\lambda-1}{\lambda}  \right)^{i\left( Z\left( q_i(x) \right)-Z\left( q_j(z) \right) \right)-1}\,.
\end{align}
where
\begin{align}\label{eq:Z(x)}
Z(x)&=\frac{1}{2\pi} \log \Omega(x)
\end{align}
with
\begin{align}\label{eq:Omegasolsupmat}
\Omega(x)
&=-\prod_{\ell} \frac{G(x,b_\ell^+)}{G(x,a_\ell^+)} \frac{G(x,g(a_\ell^-))}{G(x,g(b_\ell^-))}\,.
\end{align}

The factor $G_{ij}(x,z)$ in \eqref{eq:Fijsol} provides the pole required in \eqref{eq:Fpoles}. The function $Z$ is the result of the desired branch cut \eqref{eq:branch} and depends on both the region and the propagator. Replaced back in \eqref{eq:Fij} this gives the four components of $\F$ which in turn provides the resolvent via \eqref{eq:ansatz}. This concludes the construction of the resolvent.

A subtlety of how to obtain \eqref{eq:Fijsol} is in order. Notice the extra factor of $-1$ in the exponent of \eqref{eq:Fijsol}. The reason for its presence the following. The solution to the contour equation with $x,y\in\mathbb C$ is \eqref{eq:Fijsol} but without this factor. Then, one must take the limit $x+ i0^\pm,y+i0^\mp$ to obtain the result on $\tilde V$. In doing so, the exponent acquires an extra factor due to the cut in $\lambda$, if we wish to express now both $x$ and $y$ on the same side of the cut. Once this limit is taken, \eqref{eq:Fijsol} has no cuts along $\tilde V$, and provides the solution to the integral equation.

We finish this technical section by providing a useful formula. Replacing \eqref{eq:ansatz} back into \eqref{eq:Cauchyformula}, one finds
\begin{align}\label{eq:f(G)F}
f(\G)=\frac{1}{2\pi i} \oint_\gamma d\lambda \frac{f(\lambda)}{\lambda^2} \F
\end{align}
where we dropped the first term in \eqref{eq:ansatz} since the associated integrand is holomorphic inside $\gamma$. Then, we can use the branch cut of $\F$ in the variable $\lambda$ in order to recast the contour integral as a regular integral over the spectrum. From \eqref{eq:Fijsol} it is easy to see that $F$ just above and below the cut along $\lambda\in[0,1]$ satisfy
\begin{align}\label{}
F_{jk}(x,y|\lambda+i\epsilon)
&= \frac{\Omega(q_j(x))}{\Omega(q_k(y))} F_{jk}(x,y|\lambda-i\epsilon)\,.
\end{align}

This yields the spectral decomposition of the correlator:
\begin{align}\label{eq:spectral}
f(\G)_{jk}=\frac{1}{2\pi i} \int_0^1 d\lambda \frac{f(\lambda)}{\lambda^2} \left[ 1- \frac{\Omega(q_j(x))}{\Omega(q_k(y))} \right] F_{jk}\big|_{\lambda-i\epsilon}\,.
\end{align}

This formula is ready to be used for any desired function $f$ holomorphic in the interior of $\gamma$. The simplest application is computing the Renyi entropies, to which we now turn.

For the free fermion, it is easy to show that
\begin{align}\label{}
\log \Tr\left( \rho^n \right)=\Tr \log \left( \G^n+(1-\G)^n \right)\,.
\end{align}
To compute the entropies, we can use the spectral decomposition \eqref{eq:spectral} replacing $f=\log \left( \lambda^n+(1-\lambda)^n \right)$, taking the trace 
and performing the integral in $\lambda$. The trace is easy to compute since, due to the pole in the propagator, it yields a boundary term,
\begin{align}\label{}
\Tr \left[ \left(1- \frac{\Omega(q_\pm(x))}{\Omega(q_\pm(y))} \right) F_{\pm\pm} \right]=\pm \frac{i}{2\pi } \frac{\lambda}{\lambda-1} \log \Omega(q_\pm(x))\big|_{\partial V}\,.
\end{align} 

Replacing this into the spectral decomposition, all the $\lambda-$dependence is isolated into a prefactor given by\,\footnote{We thank Pascal Fries for a beautiful derivation of this formula.}
\begin{align}\label{}
\frac{1}{4\pi^2}\int_0^1d\lambda \frac{\log\left( \lambda^n+(1-\lambda)^n \right)}{\lambda(\lambda-1)}=\frac{1-n^2}{24n}\,.
\end{align}

The last step to obtain the entropies is to regularise their well known UV divergences. This is done by considering the region $V_\delta=\cup_{\ell}(a_\ell+\delta,b_\ell-\delta)$ with a very small $\delta>0$. Putting everything together, we arrive at the R\'enyi entropies given in the main text.

\subsection{Modular Hamiltonian}

As we saw above, the modular Hamiltonian for gaussian free fermion states takes the quadratic form $K=\int dxdy\, \psi_i^\dag(x) k_{ij}(x,y)  \psi_j(y)$, defined by the kernel $k$. Once again the resolvent techniques described before provide the answer. In this case it is given as the contour integral
\begin{align*}\label{}
k_{ij}
&=\epsilon^{\frac{i-j}{2}}\sqrt{ (-1)^{\frac{i-j}{2}} q'_iq'_j } G(q_i,q_j)\oint_\gamma d\lambda \frac{f(\lambda)}{\lambda(\lambda-1)} \left( \frac{\lambda}{\lambda-1} \right)^{i\tilde t} 
\end{align*}
where $f=-\log\left( \lambda^{-1}-1 \right)$, and we have abbreviated $q_i=q_i(x)$ and $\tilde t=Z(x)-Z(y)$. Contour integrals of this form have been evaluated in \cite{Fries:2019ozf,Erdmenger:2020nop}. The result is:
\begin{align}\label{eq:solkernel}
k_{ij}(x,y)
=-2\pi \delta \Big( Z(q_i(x))-Z(q_j(y)) \Big) G_{ij}(x,y)\,.
\end{align}

From this, two features stand out as characterising the modular Hamiltonian. First, the modular Hamiltonian couples both chiralities, since the kernel \eqref{eq:solkernel} is not diagonal: $k_{ij}$ is the coupling between chiralities $i$ and $j$.  The second feature regards its locality. By \textit{local} we mean that $k\propto \delta(x-y)$ so the modular Hamiltonian couples a point only to itself. \textit{Bi-local} means that $k\propto \delta(x-f(y))$ for some function $f$, so that $x$ is coupled only to a specific set of points $y$. \textit{Completely-nonlocal} means $k$ is a smooth function of two variables with support on $V^{2}$. As we explain in the main text, the Hamiltonian is bi-local, connecting any given point to two points in each interval. 

\section{Bulk fields on AdS$_2$}

The most direct application of our results to curved spacetimes involves two dimensional anti-de Sitter space, whose metric in the Poincar\'e patch is
\begin{align}\label{eq:AdSmetric}
ds^2=\Lambda(x)^2(-dt^2+dx^2)
\end{align}
with $\Lambda(x)=L/x$. The line $x=0$ is a (static) asymptotic conformal boundary, which light rays reach at a finite coordinate time. Thus one must specify what happens at the boundary in order to fix the physics in the bulk. 

A natural choice is to impose reflecting conditions at the asymptotic boundary. As mentioned above, a static boundary allows conformal boundary conditions, so that no energy leaks in or out of AdS. From the bulk perspective, this is indistinguishable from having the interior in a thermal state in equilibrium with a reservoir and corresponds to the analogue of the state considered in \cite{Hawking:1982dh}. 

Following the standard replica construction\,\cite{Calabrese:2004eu}, the traces of the reduced state $\rho_V$ are given by the correlator of twist operators $\mathcal{T}$ located at the endpoints:
\begin{align}\label{}
\Tr(\rho_V^n)|_{\AdS}=\prod_{\ell} \langle \mathcal{T}(a_\ell) \mathcal{\tilde T}(b_\ell) \rangle_{\AdS}\,.
\end{align}
Since the twist fields are primary with conformal weight $d_n=\frac{c}{12}\left( n-\frac{1}{n} \right)$ with $c$ the Virasoro central charge, the Weyl rescaling that takes the right half plane into $\AdS$ implies that
the R\'enyi entropies for Poincare AdS$_2$ with reflecting boundary conditions are
\begin{align}\label{eq:SAdS}
S_{\AdS}^{(n)}=S_{\text{RHP}}^{(n)}+\frac{n+1}{12n}\log \prod_\ell \frac{L^2}{a_\ell b_\ell}    
\end{align}
where the first term on the right corresponds to the entropies for the static mirror in flat space, 
and we used $c=1$ due to both chiralities.

For example, the entropy of a single interval $(a,b)$ on the AdS$_2$ vacuum is 
\begin{align}\label{SnAdSvac}
S_{\AdS}^{(n)}&=\frac{n+1}{12n}  \log \left( \frac{2L}{\delta} \frac{r-1}{r+1} \right)^2\,.
\end{align}
This depends only on the ratio $r=b/a$, given the scaling symmetry of the metric on a constant time slice. This is possible because the curvature radius of AdS compensates the dimensions of the UV cutoff $\delta$. Moreover this simple case exemplifies that the entropies of a system of infinite extension need not be divergent.

If instead of reflecting boundary conditions we simply consider two decoupled chiralities on AdS, the entropy of a single interval is determined by conformal symmetry and gives
\begin{align}\label{Sdecoupled}
\tilde S=\frac{1}{6}\log \left[ \left( \frac{L}{\delta} \right)^2 \frac{(r-1)^2}{r} \right]\ \ \ \mbox{decoupled}\,.
\end{align}

While both \eqref{SnAdSvac} and \eqref{Sdecoupled} are scale invariant, the entropies studied in the present work \eqref{SnAdSvac} are always \textit{lower} due to the monogamy of entanglement between the two chiralities. Because this arises from quantum correlations, its effect is most noticeable at low temperature. At high temperatures thermal correlations dominate and the two different states considered give perturbatively similar results. Finally, let us consider the consequences of these ideas for the problem of finding quantum extremal surfaces.



\subsection{Application: Quantum extremal surfaces}

Entanglement entropy has become a prominent topic in gauge/gravity duality mainly due to the Ryu-Takayanagi formula and its extensions\,\cite{Ryu:2006bv,Hubeny:2007xt,Faulkner:2013ana}. This is a generalisation of the Bekenstein-Hawking law, stating that the \textit{generalised} entropy of the dual field theory is given by
\begin{align}\label{eq:Sgen}
S_{\text{gen}}=\frac{A}{4G}+S_{\text{bulk}}\,.
\end{align}

Here $A$ is the area of a codimension$-2$ surface in AdS anchored at the asymptotic subregion, whereas the second term is the bulk entanglement entropy of the fields bounded between the subregion and that surface. A \textit{quantum extremal surface} (QES)\,\cite{Engelhardt:2014gca} is the minimum of the variational principle $\delta S_{\text{gen}}=0$. The results reported above find immediate applications in this context because they allow us to compute the second term of \eqref{eq:Sgen}.

We consider AdS$_2$ black hole solutions of two-dimensional Jackiw-Teiltelboim gravity\,\cite{Teitelboim:1983ux,Jackiw:1984je,Maldacena:2016upp} following the conventions of \cite{Almheiri:2019qdq}. Here, the area term in \eqref{eq:Sgen} is replaced by the dilaton field $\phi$ which measures the area of the sphere of a higher-dimensional nearly extremal black hole. The solution for the dilaton in the Poincar\'e coordinates is 
\begin{align}\label{}
\phi(x)=\phi_0+\frac{\phi_r}{x}
\end{align}
where $\phi_r$ is a positive constant and the second term comes from the area at extremality and is not important here. On top of this background, we place the fermions at zero temperature with reflecting boundary conditions. 

From the holographic perspective it is natural to take one endpoint to the conformal boundary first, $a\to \delta$, since the RT surface is anchored there, and let the endpoint on $b$ vary. However as pointed out earlier, the entropy $S_{\AdS}^{(1)}$ from \eqref{SnAdSvac} becomes a constant in this limit! Thus, the QES would require to extremise
\begin{align}\label{}
S_{\text{gen}}=\phi(b)+\frac{1}{3}\log \left( \frac{2L}{\delta} \right)+\mathcal{O}(\delta)
\end{align}
and since $\phi(x)$ is monotonic, 
\begin{align}\label{}
\partial_b S_{\text{gen}}=0\ \ \rightarrow\ \ b\to\infty\,.
\end{align}
In other words, for fermions with asymptotic reflecting boundary conditions, there is no `non-trivial' quantum extremal surface in the zero temperature solution of JT gravity, the only solution being $b\to\infty$ which corresponds to the horizon. 

Let us see how these results differ from those in encountered in recent literature\,\cite{Almheiri:2019yqk,Almheiri:2019qdq,Anegawa:2020ezn,Hollowood:2020kvk}. (Notice our conventions for the endpoints are somewhat reversed with respect to these; in order to compare we must replace $b\to a$ in our expressions, and then take (5) of \cite{Almheiri:2019yqk} in the limit $b\to 0$). These works have considered the state of two decoupled chiralities, i.e. \eqref{Sdecoupled}. Then, the associated generalised entropy $\tilde S_{\mbox{\scriptsize gen}}=\phi(x)+\tilde S$ gives
\begin{align}\label{}
\partial_b \tilde S_{\mbox{\scriptsize gen}} = 0 \ \ \ \rightarrow \ \ \ b=6\phi_r\,.
\end{align}

This is the result obtained in \cite{Almheiri:2019yqk} (again after sending the endpoint on the bath to the AdS boundary). The main difference is that have considered the case where there are quantum correlations between the ingoing and outgoing fields, $G_\pm \neq 0$, whereas these works have considered states where the two opposite chiralities are in a product state, $G_\pm=0$, but put at the same temperature so that the correlations between them are classical. 




\end{document}